\documentclass{aastex631}

\newcommand{\kms}{km~s$^{-1}$}
\newcommand{\lcou}{K~km~s$^{-1}$~pc$^2$}

\newcommand{\beq}{\begin{equation}}
\newcommand{\eeq}{\end{equation}}

\newcommand{\cm}{cm$^{-2}$}

\newcommand{\Msun}{\textrm{M}_\odot}
\newcommand{\kmps}{km~s$^{-1}$}
\newcommand{\hi}{H{\sc i}}

\newcommand{\lco}{K~km~s$^{-1}$~pc$^2$}
\newcommand{\Mmol}{\rm{M_{mol}}}
\newcommand{\cii}{[C{\textsc{ii}}]~158$\mu$m}

\begin{document}

\title{A massive H{\sc i}-absorption-selected galaxy at $z\approx2.356$ }

\author{B. Kaur} 
\affiliation{National Centre for Radio Astrophysics, Tata Institute of Fundamental Research, 
Pune University, Pune 411007, India}
\affiliation{Inter-University Centre for Astronomy and Astrophysics, Pune University, Pune 411007, India}

\author{N. Kanekar} 
\affiliation{National Centre for Radio Astrophysics, Tata Institute of Fundamental Research, Pune University, Pune 411007, India}

\author{M. Neeleman}
\affiliation{National Radio Astronomy Observatory, 520 Edgemont Road, Charlottesville, VA 22903, USA}

\author{Y. Zhu}
\affiliation{Steward Observatory, University of Arizona, 933 N Cherry Ave, Tucson, AZ 85721, USA}

\author{J. X. Prochaska}
\affiliation{Department of Astronomy \& Astrophysics, UCO/Lick Observatory, University of California, 1156 High Street, Santa Cruz, CA 95064, USA}
\affiliation{Kavli Institute for the Physics and Mathematics of the Universe (Kavli IPMU), 5-1-5 Kashiwanoha, Kashiwa, 277-8583, Japan}

\author{M. Rafelski}
\affiliation{Space Telescope Science Institute, 3700 San Martin Drive, Baltimore, MD 21218, USA}
\affiliation{Department of Physics \& Astronomy, Johns Hopkins University, Baltimore, MD 21218, USA}

\author{G. Becker}
\affiliation{University of California, Riverside, 900 University Ave, Riverside, CA 92521, USA}

\begin{abstract}
We use the Karl G. Jansky Very Large Array (VLA) and the Atacama Large Millimeter/submillimeter Array (ALMA) to detect CO(1--0), CO(3--2), and rest-frame 349-GHz continuum emission from an H{\sc i}-selected galaxy, DLA1020+2733g, at $z\approx2.3568$ in the field of the $z=2.3553$ damped Lyman-$\alpha$ absorber (DLA) towards QSO~J1020+2733. The VLA CO(1--0) detection yields a molecular gas mass of $(2.84\pm0.42)\times10^{11}\times(\alpha_{\rm CO}/4.36)\,\rm{M_\odot}$, the largest ever measured in an H{\sc i}-selected galaxy. The DLA metallicity is $+0.28 \pm 0.16$, from the 
Zn{\sc ii}$\, \lambda2026$\AA\ absorption line detected in a Keck Echellette Spectrograph and Imager spectrum. This continues the trend of high-metallicity DLAs being frequently associated with massive galaxies. We obtain a star-formation rate (SFR) of $\lesssim400$~M$_\odot$~yr$^{-1}$ from the rest-frame 349-GHz continuum emission, and a relatively long molecular gas depletion timescale of $\gtrsim0.6$~Gyr. The excitation of the J=3 rotational level is sub-thermal, with $r_{31}\equiv{L'_{\rm{CO(3-2)}}/L'_{\rm{CO(1-0)}}}=0.513\pm0.081$, suggesting that DLA1020+2733g has a low SFR surface density. The large velocity spread of the CO lines, $\approx500$~km~s$^{-1}$, and the long molecular gas depletion timescale suggest that DLA1020+2733g is likely to be a cold rotating-disk galaxy.

\end{abstract}


\section{Introduction} \label{sec:intro}

The presence of a damped Lyman$\alpha$ absorber (DLA, with \hi\ column density~$\geq 2 \times 10^{20}$~cm$^{-2}$) in a QSO absorption spectrum has long been used as a signpost of the presence of a galaxy close to or along the QSO sightline \citep[e.g.][]{Wolfe86,Wolfe05}. Numerous studies have tried to identify these galaxies at high redshifts over the last three decades \citep[e.g.,][]{Kulkarni00, Moller04, Fynbo10, Peroux12, Fumagalli14, Wang15, Krogager17}. Such galaxy samples, selected by their \hi\ absorption, are not biased towards objects with high luminosity or star-formation activity, and thus offer a view of the high-redshift Universe complementary to that provided by the usual galaxy populations identified by their stellar emission in deep optical or near-infrared images \citep[e.g.][]{Madau14}. Characterizing such high-$z$ \hi-absorption-selected galaxies and connecting them to the emission-selected population have hence long been of much interest \citep[e.g.][]{Moller02,Wolfe05,Rhodin21,Kaur21}. QSO absorption spectra also provide detailed information on the element abundance, gas-phase metallicity, molecular fraction, gas temperature, etc along the DLA sightline \citep[e.g.][]{Noterdaeme08,Rafelski14,Kanekar14, Neeleman15, deCia16, Balashev19}. This allows one to connect the properties of absorption-selected galaxies to the properties of their interstellar and circumgalactic mediums (ISMs and CGMs), not usually possible with other high-$z$ galaxy samples.

Unfortunately, despite many efforts, the presence of the nearby bright background QSO has made it difficult to even identify, let alone characterize, high-$z$ DLA galaxies via their stellar or nebular emission. Only a few tens of DLA galaxies have been identified out to $z \approx 3$ by such approaches today \citep[e.g.][]{Krogager17,Lofthouse23,Oyarzun24}, with biases implicit in the observational technique \citep[e.g. against dusty galaxies,  galaxies at high impact parameters, etc; e.g. ][]{Oyarzun24}.

The advent of the Atacama Large Millimeter/submillimeter Array (ALMA) has changed the field, opening two new windows, via \cii\ fine-structure and CO rotational lines, on high-$z$ DLA galaxies \citep[e.g.,][]{Neeleman17,Neeleman18,Neeleman19,Kanekar18, Kanekar20, Moller18, Peroux19, Klitsch19}. ALMA \cii\ searches in the fields of DLAs at $z \approx 4$ have yielded the first identifications of \hi-selected galaxies at such high redshifts \citep[][Neeleman et al.\, submitted]{Neeleman17,Neeleman19}. Remarkably, the identified DLA galaxies at $z \gtrsim 4$ are at relatively high impact parameters, $\approx 15-60$~kpc, with no evidence of a galaxy closer to the QSO sightline \citep[e.g.][]{Neeleman19,Kaur21}. This indicates that high \hi\ column densities are prevalent in the CGM of typical galaxies at $z \approx 4$. Higher-resolution ALMA \cii\ mapping of two of these galaxies has resulted in the identification of the first cold, rotating-disk galaxy, the ``Wolfe disk'', at $z > 4$ \citep{Neeleman20}, and a pair of merging galaxies at $z \approx 3.8$ \citep{Prochaska19}. Follow-up Hubble Space Telescope (HST) imaging has yielded the detection of the rest-frame near-ultraviolet stellar emission from a number of DLA  galaxies at $z \approx 4$, while Karl G. Jansky Very Large Array (VLA) CO studies have measured, or placed limits on, their molecular gas mass \citep{Neeleman20,Kaur21}. 

At somewhat lower redshifts, $z \approx 2-3$, ALMA and Northern Extended Millimetre Array (NOEMA) searches for redshifted CO emission have identified approximately 7 \hi-selected galaxies in DLA fields \citep[e.g.][]{Neeleman18,Fynbo18,Kanekar20,Kaur22c,Combes24}. While such CO studies are only sensitive to galaxies with a high molecular gas mass, it is remarkable that many high-metallicity DLAs, with [M/H]~$\gtrsim -0.3$, are associated with massive galaxies, with molecular gas mass $\gtrsim 5 \times 10^{10} \, \rm M_\odot$ \citep{Kanekar20,Kaur22b}. Evidence has been found for a dependence of both the \cii\ detection rate at $z \approx 4$ and the CO detection
rate at $z \approx 2$ on DLA metallicity, with higher-metallicity DLAs at each redshift yielding a significantly higher detection rate of galaxies in the DLA field. Recently, \citet{Kaur24} carried out the first CO mapping of a high-$z$ \hi-selected galaxy, finding evidence that the $z \approx 2.193$ DLA towards PKS~B1228-113 also arises in a massive, cold, dusty rotating disk.

Only $\approx 20$~DLA fields have so far been searched for redshifted CO emission at $z \approx 2$ \citep{Kanekar20,Kaur22c}. The efficiency of such searches can be significantly increased by targetting sightlines containing multiple DLAs, such that multiple DLA redshifts can be simultaneously searched for CO emission. We have begun such a programme to use ALMA, NOEMA, and the VLA to search for redshifted CO emission from such ``double DLA'' fields at $z \approx 2$. Here, we report the ALMA and VLA detection of redshifted CO emission from an ultra-massive \hi-selected galaxy at $z \approx 2.356$, from the double-DLA field towards QSO~J1020+2733.\footnote{We use a flat $\Lambda$ cold dark matter cosmology, with $\Omega_{\Lambda} = 0.7$, $\Omega_{m} = 0.3$, and H$_0 = 70$~\kmps~Mpc$^{-1}$, throughout the paper.}

\section{Observations and Data Analysis} 
\label{sec:obs}

The sightline towards QSO~J1020+2733 (at $z = 2.712$) was identified as containing two DLAs, at redshifts of $z \approx 2.36$ and $z \approx 2.29$ in a Sloan Digital Sky Survey (SDSS)  spectrum \citep{Mas-Ribas17,Arinyo18}. The redshifted CO lines of both DLAs were simultaneously covered in the ALMA CO(3--2) and the VLA CO(1--0) observations. We also used the Echellette Spectrograph and Imager \citep[ESI; ][]{Shienis02} onboard the Keck-II Telescope to obtain an optical spectrum of QSO~J1020+2733, to accurately measure the metallicities of the two putative DLAs. Finally, we re-analysed the SDSS spectrum of QSO~J1020+2733, to provide updated \hi\ column densities of the two absorbers, based on the more accurate redshift information from the metal lines detected in the ESI spectrum.

\subsection{VLA observations and data analysis}

The VLA observations (proposal ID:~VLA/23B-286; PI: N.~Kanekar) were carried out on 30 and 31~December, 2023, using the Ka-band receivers in the D-configuration, with a total on-source time of $\approx 3$h. The WIDAR correlator was set up in 8-bit mode, with two 1~GHz intermediate frequency (IF) bands, each divided into eight 128-MHz digital sub-bands. Two of the sub-bands were centred at the redshifted CO(1--0) line frequencies for the DLA redshifts (35.06~GHz and 34.34~GHz), with 256~channels, a velocity resolution of $\approx 4.3$~\kmps, and a velocity coverage of $\approx 1100$~\kmps. The remaining sub-bands were placed at contiguous frequencies, covering the frequency range $34.62 - 35.64$~GHz, with a velocity resolution of $\approx 8.5$~\kmps.

The VLA data were analysed following standard procedures in the Common Astronomy Software Application \citep[{\sc casa} version~5.6.1.8; ][]{casa22}. After initial data editing, the tasks {\sc gaincalR} and {\sc bandpassR}, from the {\sc calR} package\footnote{The {\sc calR} package is publicly available at \url{https://github.com/chowdhuryaditya/calR} \citep{calR,Chowdhury20,Chowdhury22b}.}were used to measure the antenna-based gains and bandpasses for each observing run. These were applied
to the multi-channel visibility data and the calibrated target visibilities from the two runs were then split out and combined together into a single data set. The spectral windows containing the two redshifted CO(1--0) line frequencies were then imaged using the task {\sc tclean}, with natural weighting and a velocity resolution of 50~\kmps, to obtain the final spectral cubes, with an angular resolution of $\approx 3\farcs0 \times 2\farcs4$. The cubes were cleaned down to a flux density threshold of $0.5\sigma$, using a mask of size $\approx 12'' \times 12''$ which encloses all the CO(1--0) emission, where $\sigma \approx 0.14$~mJy~Beam$^{-1}$ is the root mean square (RMS) noise per 50~\kmps\ channel on the cubes.

\subsection{ALMA observations and data analysis}

The ALMA observations (proposal ID: 2023.1.01415; PI: J.~X.~Prochaska) were carried out on 12~March, 2024, using 45 ALMA antennas in the C43-1 configuration, the Band-3 receivers, and a total on-source time of $\approx 48$m. The ALMA correlator was set up in 4-bit Frequency Division Mode (FDM) for the three intermediate frequency (IF) bands covering the redshifted CO(3--2) line frequency for $z = 2.2876$ and  $z = 2.3564$ (i.e. frequencies of $\approx 105.18$~GHz and $\approx 103.03$~GHz, respectively), and in Time Division Mode (TDM) for the fourth band, used to cover the continuum emission. The three FDM bands each had a bandwidth of 937.5~MHz, subdivided into 120~channels, i.e. a velocity coverage of $\approx 2700$~\kmps, and a velocity resolution of $\approx 22.5$~\kmps, at the redshifted CO(3--2) line frequencies, while the TDM band used a bandwidth of 2~GHz and 128~channels.

The initial data editing and calibration of the ALMA data used the standard ALMA pipeline in the {\sc casa} package \citep[version~6.5.4.9;][]{casa22,Hunter23}. We then used manual flagging to remove outliers in the calibrated visibilities, before averaging the data to a coarse spectral resolution for the purpose of continuum imaging. The imaging used Briggs weighting \citep{Briggs95} with a robust value of $+1$, yielding a synthesized beam of $3\farcs 7 \times 3\farcs 2$ and an RMS noise of 27~$\mu$Jy~Beam$^{-1}$ on the 104~GHz continuum image. We then used the task {\sc uvsub} to subtract out the continuum emission from the calibrated visibilities. Finally, we used the task {\sc tclean} to make spectral cubes for the IF bands covering the redshifted CO(3--2) emission from the two DLA redshifts, using Briggs weighting with a robust value of +1 and a velocity resolution of 23~\kmps. The cubes were cleaned down to a flux density threshold of $0.5\sigma$, using a mask of size $\approx 15'' \times 15''$ centered on the CO emission, where $\sigma = 0.37$~mJy~Beam$^{-1}$. The angular resolution of the spectral cubes is $\approx 3\farcs7 \times 3\farcs1$.

\subsection{Keck ESI observations and data analysis}

QSO~J1020+2733 was observed on June~1, 2024 using ESI on the Keck-II telescope in echellette mode, with a 1\arcsec\ slit and a total exposure time of 1,200~seconds. The data were analysed following the reduction procedures outlined in \citet{Becker19}, using a custom pipeline. This pipeline includes optimal sky subtraction techniques \citep{Kelson03}, one-dimensional spectral extraction \citep{Horne86}, and telluric absorption corrections, which were based on atmospheric models from the Cerro Paranal Advanced Sky Model \citep{Noll12,Jones13}. The spectra were extracted with a pixel size of 15~\kmps; the typical velocity resolution is approximately 45~\kmps\ (FWHM) and the S/N per pixel redward of the Ly$\alpha$ emission line of the quasar is $\approx 10$.

\begin{figure}[!t]
\centering
\includegraphics[width = 0.5\textwidth]{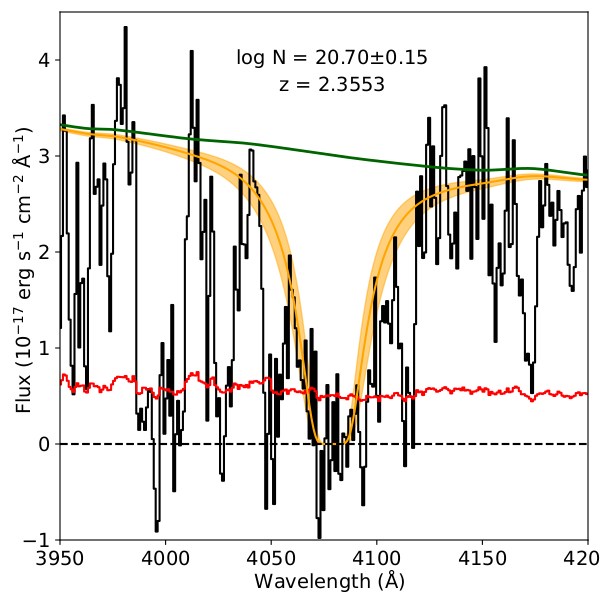}
\caption{The SDSS spectrum of QSO~J1020+2733 centred on the Ly$\alpha$ absorption feature at $z=2.3553$. The green line is the continuum fit to the quasar spectrum obtained with the python package \textit{linetools}. The orange line is a fit to the Ly$\alpha$ absorption profile at $z=2.3553$, while the orange-shaded region shows the 1$\sigma$ uncertainty on the fit. The 1$\sigma$ uncertainty on the spectrum is shown in red. The spectrum also shows saturated Ly$\alpha$ absorption at $z=2.2876$ ($\approx4000$~\AA).
\label{fig:dlafit}}
\end{figure}

\begin{figure}[!t]
\centering
\includegraphics[width = 0.5\textwidth]{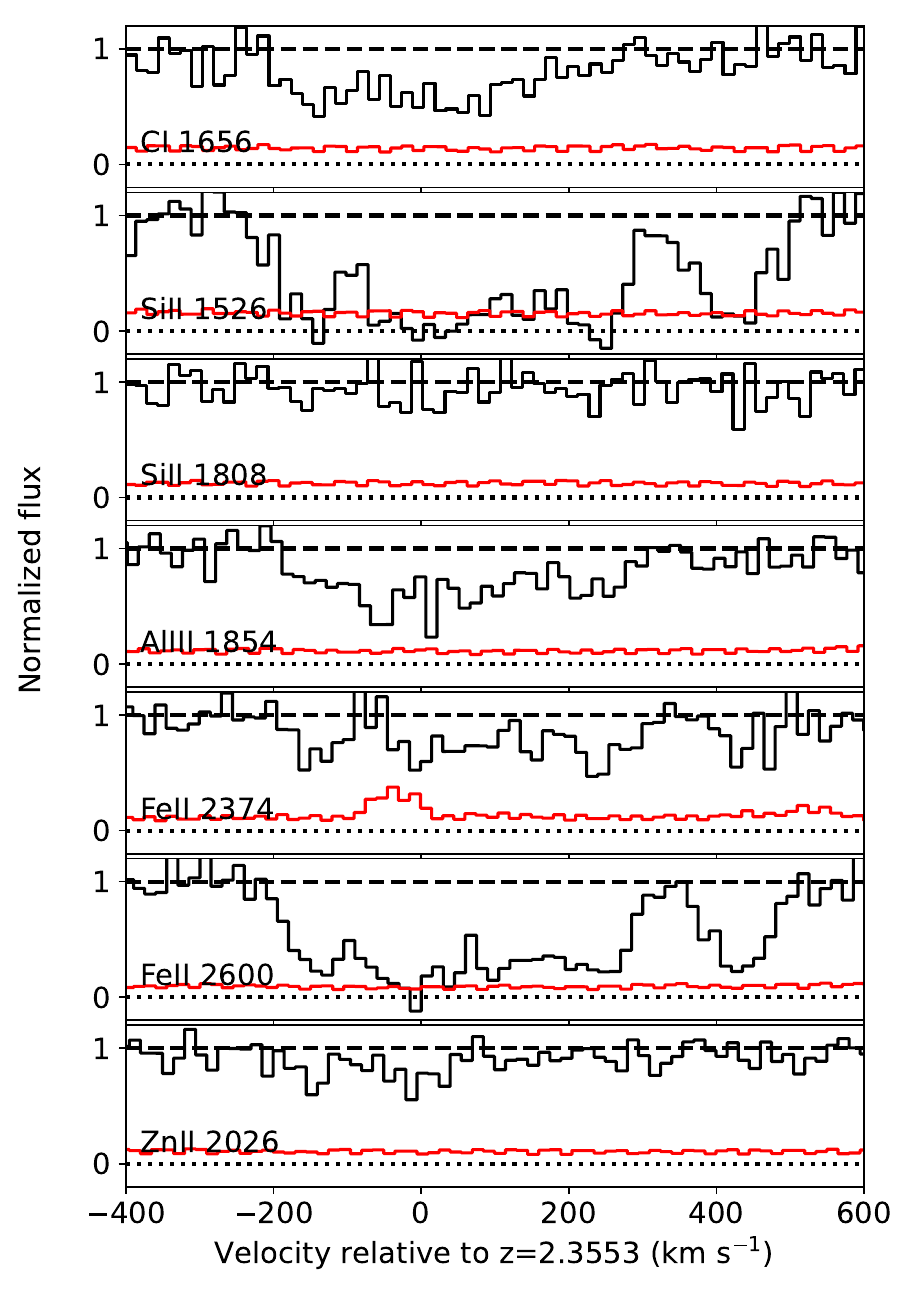}
\caption{A selection of low- and intermediate-ion metal lines associated with the DLA at $z=2.3553$ towards QSO~J1020+2733, detected in the ESI spectrum. The 1$\sigma$ uncertainty on the ESI spectrum is shown in red. The metal column densities determined from the detected lines, over the velocity range (-350, +500)~\kms, are listed in Table~\ref{table:abslines}. Absorption is seen over a velocity range of $\approx 700$~\kms, with $\Delta v_{90} = 480 \pm 50$~\kms, measured from the Zn{\sc ii}$\lambda$2026 transition.
\label{fig:abslines}}
\end{figure}

\begin{figure}[!t]
\includegraphics[width = \textwidth]{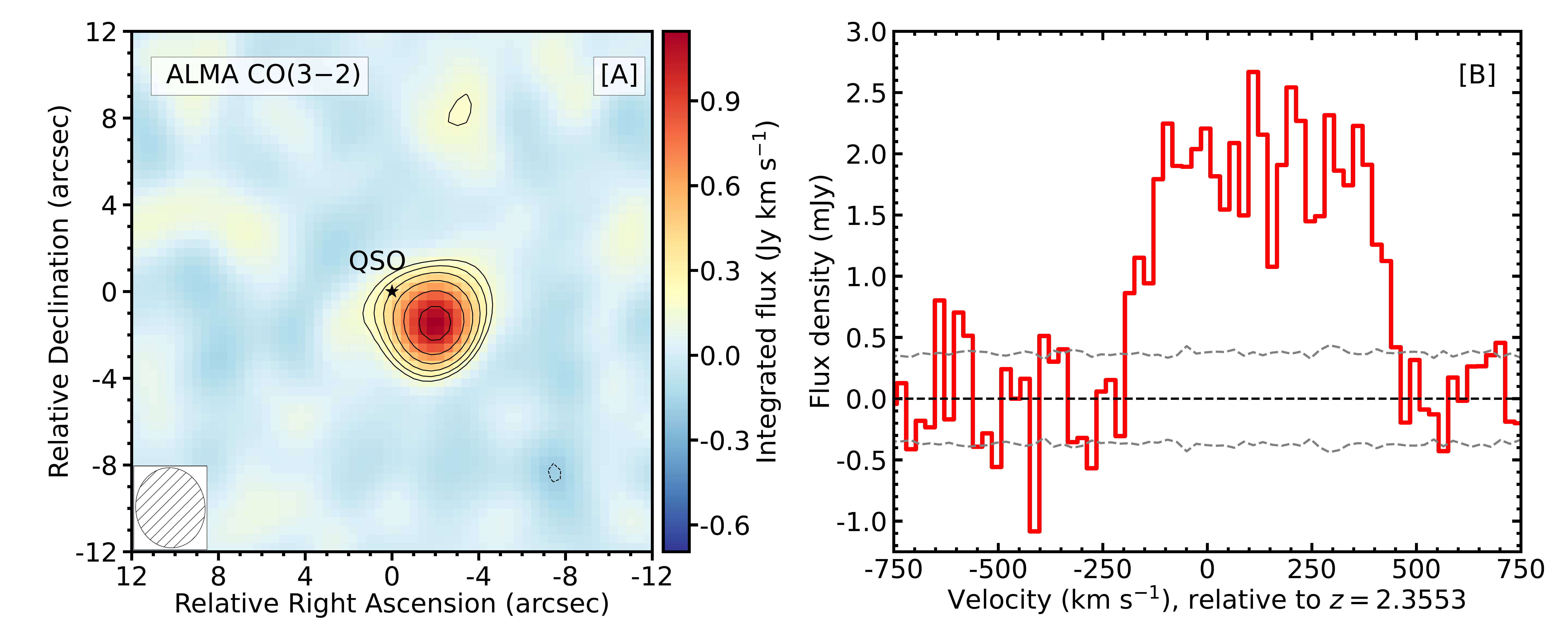}
\caption{ALMA CO(3--2) emission from the \hi-selected galaxy in the field of the $z = 2.3553$ DLA towards QSO~J1020+2733. [A]~The velocity-integrated ALMA CO(3--2) image of DLA1020+2733g. The contours are at $(-3.0, 3.0, 4.2, 6, 8.5, 12, 17.0) \times \sigma$ significance, with dashed negative contours. The axes are relative to the QSO position, indicated by the black star. [B]~The ALMA CO(3--2) emission spectrum of DLA1020+2733g, obtained by taking a cut through the CO cube at the location of the peak in [A]. The dashed grey curves indicate the $\pm1\sigma$ error on the spectrum.
\label{fig:alma}}
\end{figure}

\begin{figure}[ht]
\centering
\includegraphics[width = 0.45\textwidth]{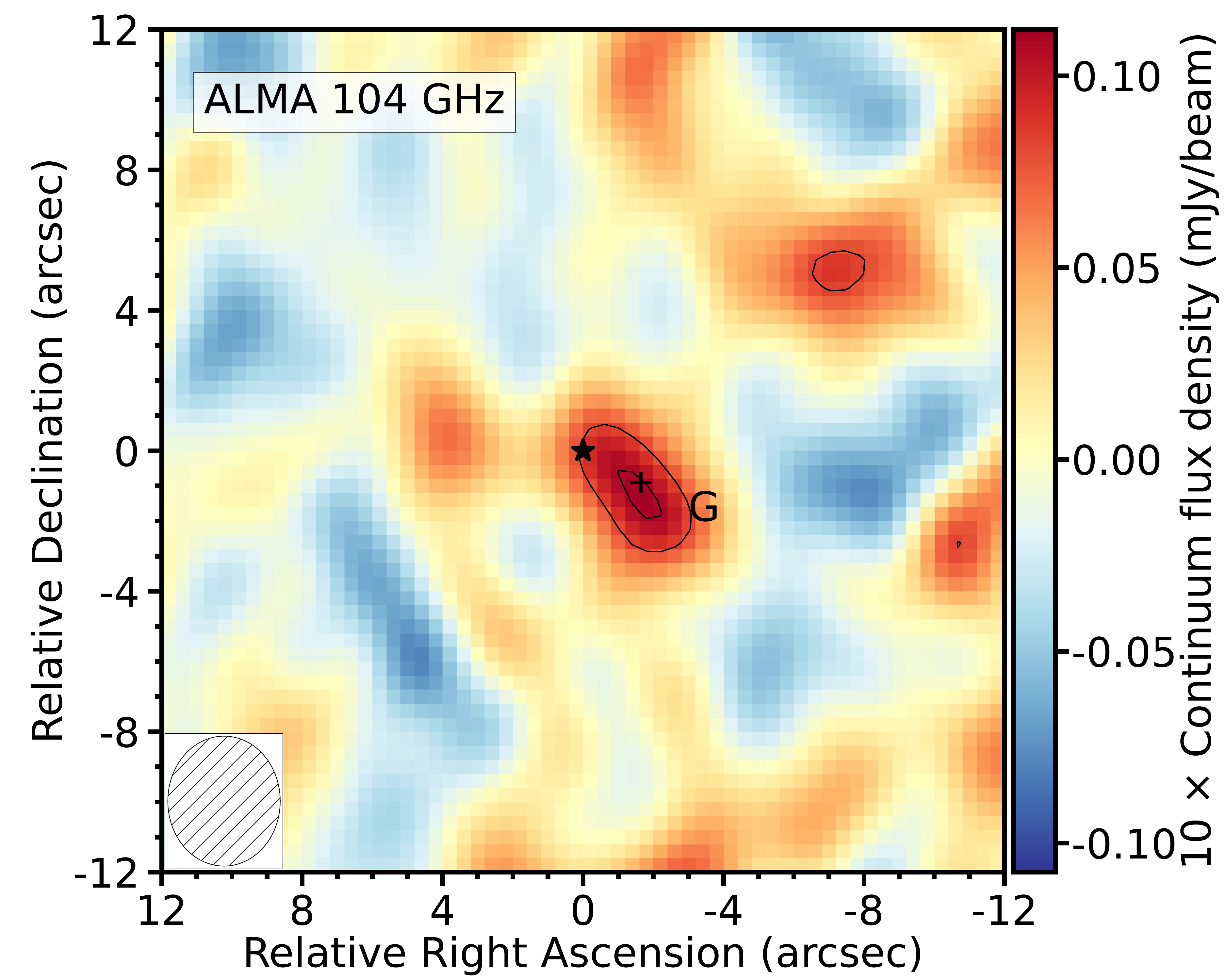}
\caption{ALMA 104~GHz continuum image of the field of the $z = 2.3553$ DLA towards QSO~J1020+2733. The location of the \hi-selected galaxy DLA1020+2733g is indicated by the black plus, marked as ``G". The contours are at $(3.0, 4.0) \times \sigma$ significance, where $\sigma = 27\mu$Jy is the RMS noise on the continuum image. We note that there are no negative contours at $\leq -3\sigma$ significance in the image. The axes are relative to the QSO position, indicated by the black star.
\label{fig:almacont}}
\end{figure}

\section{Results} 
\label{sec:results}

\begin{figure}[!t]
\includegraphics[width = \textwidth]{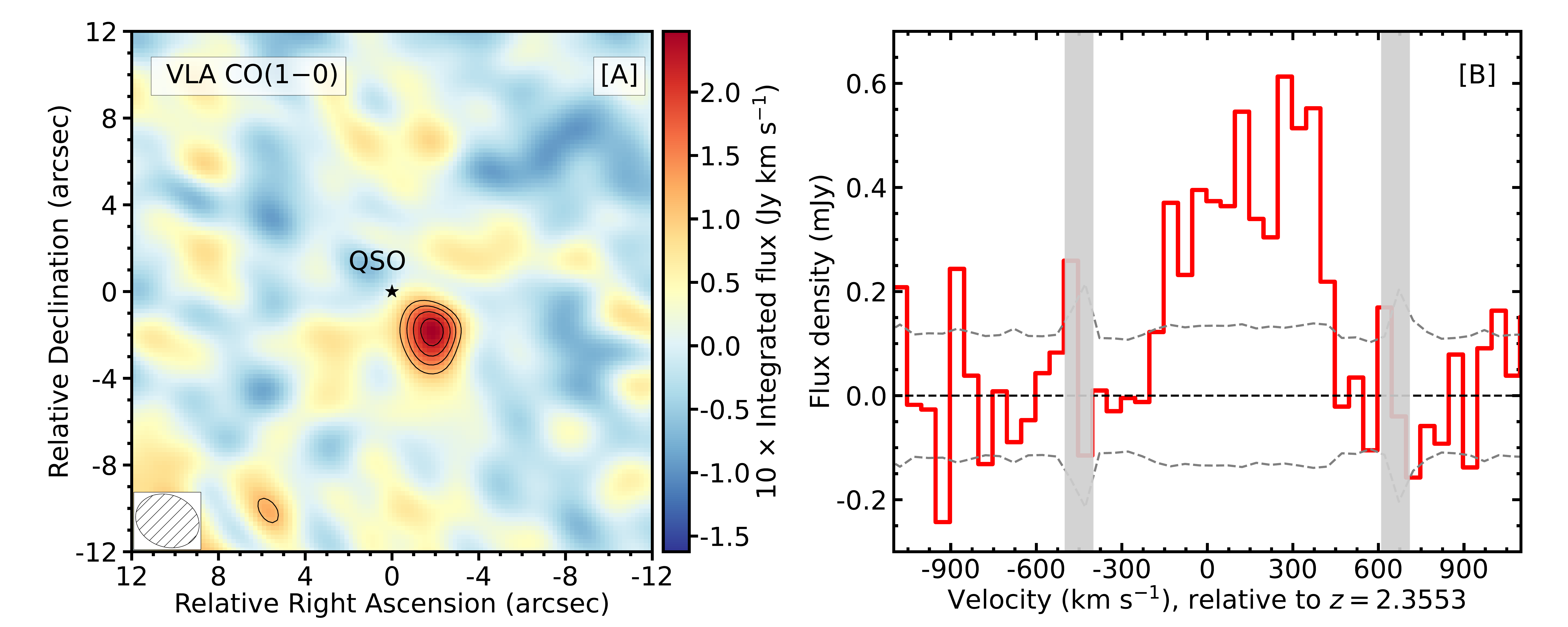}
\caption{VLA CO(1--0) emission from the \hi-selected galaxy in the field of the $z = 2.3553$ DLA towards QSO~J1020+2733. [A]~The velocity-integrated VLA CO(1--0) image of DLA1020+2733g. The contours are at $(3.0, 4.0, 5.0, 6.0) \times \sigma$ significance. We note that there are no negative contours at $\leq -3\sigma$ significance in the image. The axes are relative to the QSO position, indicated by the black star. [B]~The VLA CO(1--0) emission spectrum of DLA1020+2733g, obtained by taking a cut through the CO cube at the location of the peak in [A]. The grey bands represent the edges of the digital subbands of the WIDAR correlator. The dashed grey curves indicate the $\pm1\sigma$ error on the spectrum.
\label{fig:vla}}
\end{figure}

\begin{table}[!b]
\centering
\caption{Properties of the $z = 2.3553$ DLA. The metallicity [M/H] is based on the Zn{\sc ii}$\lambda$2026\AA\ line. Note that we have not included an estimate of the C{\sc i} column density, as the C{\sc i}$\lambda$1656\AA\ line is blended with a complex of C{\sc i}* and C{\sc i}** lines at the ESI spectral resolution.
\label{table:abslines}}
\begin{tabular}{ll}
\hline
\multicolumn{2}{c}{DLA~J1020+2733}\\
\hline
\hline
Right Ascension (J2000) & 10:20:29.88 \\
Declination (J2000) & +27:33:28.66\\
Redshift & 2.3553\\
$\log(N$(H\,\textsc{i})/cm$^{-2})$ & $20.70 \pm 0.15$\\
$\rm{[M/H]}$ & $+0.28 \pm 0.16$\\
$\Delta V_{90}$ (\kms) & $480 \pm 50$\\
$\log(N$(C\,\textsc{iv})/cm$^{-2})$ & $> 15.37$\\
% $\log(N$(C\,\textsc{i})/cm$^{-2})$ & $ 14.56 \pm 0.04$\\
$\log(N$(Al\,\textsc{ii})/cm$^{-2})$ & $> 14.16$\\
$\log(N$(Al\,\textsc{iii})/cm$^{-2})$ & $13.96 \pm 0.03$\\
$\log(N$(Si\,\textsc{ii})/cm$^{-2})$ & $15.79 \pm 0.11$\\
$\log(N$(Si\,\textsc{iv})/cm$^{-2})$ & $> 14.78$\\
$\log(N$(Fe\,\textsc{ii})/cm$^{-2})$& $14.95 \pm 0.02$\\
$\log(N$(Zn\,\textsc{ii})/cm$^{-2})$ & $13.59 \pm 0.06$\\
\hline
\end{tabular}
\end{table}

Our Keck-ESI spectrum of QSO~J1020+2733 revealed a slew of metal lines at $z = 2.3553$, but no strong metal absorption was seen at the redshift of the second SDSS DLA at $z = 2.2876$. While saturated Ly$\alpha$ absorption is indeed seen at $z = 2.2876$ in the SDSS spectrum of QSO~J1020+2733, there is no evidence of damping wings. This absorber is thus likely to be a blend of Lyman-limit systems, and not a DLA, and we will not discuss it further here.

Figure~\ref{fig:dlafit} shows a zoom-in on the damped Ly$\alpha$ absorption detected at $z = 2.3553$ towards QSO~J1020+2733 in the SDSS spectrum\footnote{We note that the DLA lies at the extreme blue end of the ESI spectrum, where the S/N is very low; we have hence measured the \hi\ column density from the SDSS spectrum.}. The solid orange curve shows our fit to the damped Ly$\alpha$ profile, while the orange-shaded region shows the $1\sigma$ uncertainty on the fit. From the fit to the \hi\ absorption profile, and in particular the observed damping wings \citep[e.g.,][]{Wolfe05,Draine11}, we obtain an \hi\ column density of $(5.0 \pm 1.7) \times 10^{20}$~\cm, i.e. log[$N_{\rm HI} / {\rm cm}^{-2}] = 20.70 \pm 0.15$. 

The ESI spectrum shows absorption at $z = 2.3553$ from a wide range of metal lines (see Fig.~\ref{fig:abslines}). We used the apparent optical depth method to measure the metal column densities and element abundances for this DLA; these are listed in Table~\ref{table:abslines}. We obtain a DLA metallicity of [M/H]~$=+0.28 \pm 0.16$ from the Zn{\sc ii}$\lambda$2026\AA\ absorption line. The uncertainty on this determination is largely driven by uncertainties in the quasar continuum placement in the \hi\ column density estimate. We also obtain high Fe and Si abundances ([Fe/H]~=~$-0.06 \pm 0.15$ and [Si/H]~=~$-0.13 \pm 0.18$), consistent with a super-solar metallicity. We note that, curiously, the dust depletion is only [Zn/Fe]~$\approx +0.34$, relatively low for DLAs at such a high metallicity \citep[e.g.][]{Rafelski12,deCia16}. The velocity spread of the metal lines is $\Delta V_{\rm 90} = (480 \pm 50)$~\kms, also from the Zn{\sc ii}$\lambda$2026\AA\ line; this is one of the largest $\Delta V_{\rm 90}$ values ever measured in a DLA \citep[e.g.][]{Neeleman13}.

%However, we note that this metallicity estimate is consistent with the abundance of Fe ([Fe/H] = $-0.06 \pm 0.15$) assuming a significant amount of depletion of Fe on to dust grains, as is typical for DLAs at such high metallicities \citep[e.g.][]{Rafelski12}. The velocity spread of the metal lines is $\Delta V_{\rm 90} = (480 \pm 50)$~\kms, also from the Zn{\sc ii}$\lambda$2026\AA\ line.

Our ALMA spectral cube revealed a high-significance detection of CO(3--2) emission in the field of the $z = 2.3553$ DLA. Fig.~\ref{fig:alma}[A] shows the velocity-integrated ALMA CO(3--2) image of the field. The CO(3--2) emission peak is at RA=10h20m29.76s, Dec.=+27$^\circ33'27.20''$. While the emission appears to be marginally spatially resolved along the east-west direction, the integrated flux density obtained by fitting a 2-D Gaussian to the emission region is consistent with the peak flux density. The \hi-selected galaxy, hereafter referred to as DLA1020+2733g, lies at an impact parameter of $\approx 1.96''\pm0.02''$, i.e. $16.1\pm0.2$~kpc at $z = 2.3553$, south-west of the QSO location. 

Figure~\ref{fig:alma}[B] shows the ALMA CO(3--2) spectrum, obtained by taking a cut through the CO cube at the location of the peak in 
%integrating over the region showing emission in 
the moment-0 image of Fig.~\ref{fig:alma}[A]. The CO(3--2) emission has a large FWHM, $\approx 500$~\kmps.  The galaxy has a redshift of $z_{\rm gal} = 2.3568$, obtained from a single-Gaussian fit to the CO(3--2) spectrum of Fig.~\ref{fig:alma}[B]; we will use this as the galaxy redshift in the subsequent analysis. We obtain a velocity-integrated CO(3--2) line flux density of $(1.145 \pm 0.059)$~Jy~km~s$^{-1}$, implying a CO(3--2) line luminosity of $L'_{\rm CO(3-2)} = (3.34 \pm 0.17) \times 10^{10}$~\lcou.

Our ALMA 104~GHz continuum image of the field of QSO~J1020+2733 shows clear emission coincident with DLA1020+2733g (see Fig.~\ref{fig:almacont}), with a peak flux density of $(112\pm29)~\mu$Jy. The continuum emission from the \hi-selected galaxy appears extended in the direction of the QSO location. The low impact parameter, $\approx 16$~kpc, of the galaxy implies that we cannot rule out the possibility that the extended continuum emission is due to blending of the emission from the galaxy and the QSO. However, we also cannot formally rule out the possibility that the \hi-selected galaxy is solely responsible for the observed extended continuum emission. To measure the flux density of the galaxy, we fitted the continuum emission with a 1-component 2-D Gaussian model, using the {\sc casa} task {\sc imfit}. This yielded a 104~GHz continuum flux density of $(128\pm29)~\mu$Jy for DLA1020+2733g. The location of the peak of the continuum emission obtained from the Gaussian fit (indicated by the plus sign in Fig.~\ref{fig:almacont}) is in agreement with that of the CO(3--2) emission peak. We note that the measured 104~GHz continuum flux density should be considered an upper limit to the flux density of DLA1020+2733g, as some of the emission may arise from the QSO.

Finally, our VLA spectral cube also yielded a clear detection of CO(1--0) emission from DLA1020+2733g. Figure~\ref{fig:vla}[A] shows the velocity-integrated VLA CO(1--0) image of the galaxy, while Fig.~\ref{fig:vla}[B] shows the spectrum obtained by taking a cut through the VLA cube at the peak of the CO(1--0) moment-0 image. The velocity spread of the CO(1--0) line is $\approx 500$~\kms, in excellent agreement with that of the CO(3--2) line. The velocity-integrated CO(1--0) line flux density is $(0.248 \pm 0.037)$~Jy~\kms, yielding a CO(1--0) line luminosity of $L'_{\rm CO(1-0)} = (6.51 \pm 0.97) \times 10^{10}$~\lcou. No continuum emission was detected from the DLA galaxy in the VLA 35~GHz continuum image. Table~\ref{table:emlines} lists the emission properties of DLA1020+2733g.

\begin{figure}[!t]
\centering
\includegraphics[width = 0.5\textwidth]{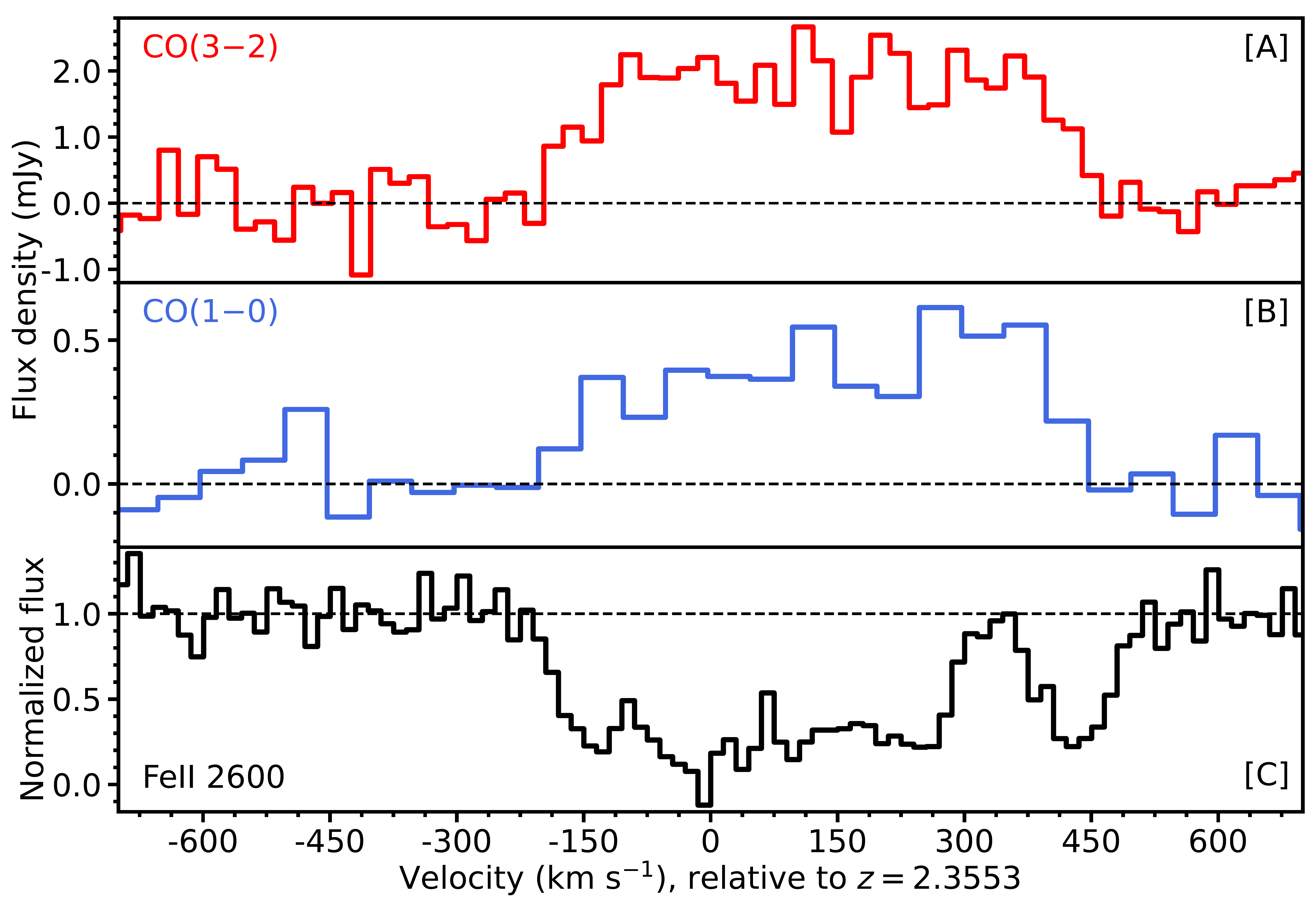}
\caption{A comparison between the velocity profiles of [A]~the CO(3--2) emission and [B]~the CO(1--0) emission detected from DLA1020+2733g, with that of [C]~the low-ionization metal absorption line, Fe{\sc ii}$\lambda$2600\AA. The velocity scale is relative to the DLA redshift, $z=2.3553$. The velocity extent of the absorption and emission are seen to be in remarkable agreement.
\label{fig:spcCompare}}
\end{figure}

\begin{table}[!b]
\centering
\caption{Emission properties of the \hi-absorption-selected galaxy, DLA1020+2733g. The estimates of the molecular gas mass and the molecular gas depletion timescale assume $\alpha_{\rm CO} \approx 4.36 \, \rm M_\odot$~(\lcou)$^{-1}$.
\label{table:emlines}}
\begin{tabular}{ll}
\hline
\multicolumn{2}{c}{DLA1020+2733g}\\
\hline
\hline
Right Ascension (J2000) & 10:20:29.775 \\
Declination (J2000) & +27:33:27.28\\
Redshift & 2.3568\\
Impact parameter (kpc) & $16.1\pm0.2$\\
$L'_{\rm CO(3-2)}$ (\lcou) & $(3.34 \pm 0.17) \times 10^{10}$\\
$L'_{\rm CO(1-0)}$ (\lcou) & $(6.51 \pm 0.97) \times 10^{10}$\\
S$_{\rm cont}$ ($\mu$Jy), at $\nu_{\rm obs}$=104 GHz  & $(112\pm29)$ \\
$\Mmol \, (\Msun)$ & $(2.84 \pm 0.42) \times 10^{11}$ \\
SFR ($\Msun~\rm yr^{-1}$)& $\lesssim400$ \\
t$_{\rm dep}$ (Gyr) & $\gtrsim0.6$ \\
$r_{31}$ & $(0.513\pm0.081)$ \\
\hline
\end{tabular}
\end{table}

\section{Discussion} \label{sec:dis}

Our VLA detection of CO(1--0) emission allows a direct estimate of the molecular gas mass of DLA1020+2733g, without any assumptions about the CO excitation. Assuming a CO-to-H$_2$ conversion factor of $\alpha_{\rm CO} = 4.36 \, \rm M_\odot$~(\lcou)$^{-1}$, applicable to main-sequence galaxies of solar metallicity \citep[e.g.][]{Bolatto13,Tacconi20}, we obtain $\Mmol = (2.84 \pm 0.42) \times (\alpha_{\rm CO}/4.36) \times 10^{11} \, \Msun$. This is the largest molecular gas mass ever measured in an \hi-selected galaxy \citep[e.g.][]{Peroux19,Kanekar18, Kanekar20,Kaur22b}. 

The measured 104~GHz (i.e. rest-frame $\approx 349$~GHz) flux density of DLA1020+2733g can be used to determine the total SFR of the \hi-selected galaxy. For this, we assume a dust temperature of $\approx 35$~K and an emissivity of $\beta = 1.5$, and then fit a modified black-body spectrum to the measured rest-frame 349~GHz flux density \citep[e.g.][]{Neeleman18}. Using the integrated continuum flux density of S$_{104~\rm GHz} = 128~\mu$Jy yields a  total infrared (TIR) luminosity between 8~$\mu$m and 1000~$\mu$m of $\approx 3 \times 10^{12}~\rm L_\odot$. Using the correlation between the TIR luminosity and the SFR \citep{Kennicutt12} gives an SFR of $\approx 400\, \rm M_\odot$~yr$^{-1}$ for DLA1020+2733g. The uncertainty on the SFR estimate is high, as it is based on continuum flux densities measured at a rest-frame frequency of $\approx 349$~GHz, far away from the  peak frequency of the dust continuum emission, and thus depends strongly on the assumed values of $\beta$ and the dust temperature. Further, we cannot formally rule out the possibility that the detected 104~GHz continuum emission contains a contribution from the QSO. Our SFR estimate from the 104~GHz image should hence be considered an upper limit to the SFR of DLA1020+2733g.

The ratio of the above estimates of the molecular gas mass and the SFR yields a molecular gas depletion timescale of $\tau_{\rm dep} \gtrsim0.6$~Gyr. This is higher than the typical molecular gas depletion timescale of galaxies on the main sequence at $z \gtrsim 2$ \citep[$\approx 0.2$~Gyr; ][]{Tacconi20}, indicating that DLA1020+2733g has a relatively low star-formation efficiency for its molecular gas mass. We note that similarly high gas depletion timescales have been found in three other \hi-selected galaxies at $z > 2$, DLA1228-113g at $z \approx 2.193$ \citep[$\approx 1.1$~Gyr; ][]{Neeleman18,Kaur22b},  DLA0918+1636g at $z \approx 2.5832$ \citep[$\approx 0.5$~Gyr; ][]{Fynbo18,Kaur22b}, and DLA0817+1351g at $z \approx 4.258$ \citep[$\approx 0.8$~Gyr; ][]{Neeleman20}. This suggests that selection by \hi-absorption tends to pick out gas-rich galaxies with lower star formation efficiencies than typical main-sequence galaxies at similar redshifts, perhaps due to the lack of bias towards high star-formation activity. However, as noted above, the uncertainties in these SFR estimates are fairly high as they are based on flux densities in the long wavelength tail of the dust continuum emission. %This suggests that selection by \hi-absorption tends to pick out more quiescent galaxies, perhaps due to the lack of bias towards high star-formation activity. However, as noted above, the uncertainties in these SFR estimates are fairly high as they are based on flux densities in the long wavelength tail of the dust continuum emission.

The above estimates of molecular gas mass and depletion timescale assume $\alpha_{\rm CO} = 4.36 \, \rm M_\odot$~(\lco)$^{-1}$. If DLA1020+2733g has a lower CO-to-H$_2$ conversion factor, $\alpha_{\rm CO} \approx 1 \, \rm M_\odot$~(\lcou)$^{-1}$, the molecular gas mass and depletion timescale would each be lower by a factor of $\approx 4.5$. However, such low $\alpha_{\rm CO}$ values are applicable in starburst galaxies, which have far shorter molecular gas depletion timescales, $< 0.1$~Gyr \citep[e.g.][]{Tacconi20}, than main-sequence galaxies. For $\alpha_{\rm CO} \approx 1 \, \rm M_\odot$~(\lcou)$^{-1}$, we would obtain $\tau_{\rm dep} \approx 0.14$~Gyr, similar to that of main-sequence galaxies at $z \approx 2.3$ \citep[e.g.][]{Tacconi20}. This relatively long molecular gas depletion timescale makes it unlikely that DLA1020+2733g is a starburst galaxy, with a low value of $\alpha_{\rm CO}$. We hence conclude that $\alpha_{\rm CO} \approx 4.36 \, \rm M_\odot$~(\lcou)$^{-1}$ is likely to apply to the $z = 2.3568$ \hi-selected galaxy, and that our estimates of $\Mmol = (2.84 \pm 0.42) \times 10^{11} \, \Msun$ and $\tau_{\rm dep} \approx 0.6$~Gyr are likely to be reliable. In passing, using a metallicity-dependent CO-to-H$_2$ conversion factor and the DLA metallicity of $+0.28 \pm 0.16$ yields $\alpha_{\rm CO} \approx 3.5 \, \rm M_\odot$~(\lco)$^{-1}$ \citep[e.g. ][]{Bolatto13}. Using this value of $\alpha_{\rm CO}$, the inferred molecular gas mass is consistent with our estimate based on $\alpha_{\rm CO} = 4.36 \, \rm M_\odot$~(\lco)$^{-1}$. 

Figure~\ref{fig:spcCompare} shows a comparison between the velocity profiles of the CO(3--2) emission, the CO(1--0) emission, and a low-ionization metal absorption line (Fe{\sc ii}$ \lambda2600$\AA) identified in the ESI spectrum. We note that the absorption arises at a transverse distance of $\approx16$~kpc from DLA1020+2733g. Given this, it is remarkable that the absorption and emission lines show very similar velocity extents, extending over $\approx 700$~\kms. However, the centroids of the CO emission from DLA1020+2733g are offset from the centroid of the Fe{\sc ii}$\lambda$2600\AA\ line by $\approx134$~\kms. 

The ratio of the luminosities in the CO(3--2) and CO(1--0) lines allows us to determine the excitation of the J=3 rotational level, $r_{31}$. We obtain $r_{31} \equiv L'_{\rm CO(3-2)}/L'_{\rm CO(1-0)} = 0.513 \pm 0.081$, indicating sub-thermal excitation of the J=3 level \citep[albeit a higher excitation than in the disk of the Milky Way; e.g.][]{Fixsen99}. Our estimate is in excellent agreement with the median value of $r_{31} \approx 0.55$ in galaxies near the main sequence at $z \approx 0-3$ \citep[e.g.][]{Daddi15,Tacconi20}. Earlier studies of massive \hi-selected galaxies, as well as massive main-sequence galaxies, at $z \gtrsim 2$ have yielded near-thermal excitation of the J=3 level, $r_{31} \approx 1$ \citep{Bolatto15,Henriquez-Brocal22,Kaur22b}. This is the first case of definite sub-thermal excitation of the J=3 level in an \hi-selected galaxy at $z > 2$. \citet{Kaur22b} noted that the near-thermal excitation of the J=3 rotational level in massive galaxies at $z > 2$ appears to be associated with a high SFR surface density in these galaxies \citep[e.g.][]{Daddi15, Boogaard20}, as also suggested by theoretical models \citep[e.g.][]{Narayanan14}. This suggests that DLA1020+2733g may have a relatively extended molecular disk, yielding a low SFR surface density and hence a lower excitation of the mid-J rotational levels. 

The large velocity FWHMs ($\approx 500$~\kms) of the CO(1--0) and CO(3--2) lines in DLA1020+2733g are similar to the FWHMs ($\approx 600$~\kms) of these lines in the \hi-selected galaxy DLA1228-113g at $z \approx 2.193$ \citep{Neeleman18,Kaur22b}. Both galaxies also have large molecular gas masses and high gas depletion timescales, suggesting that they are relatively gas-rich systems, with a low star-formation efficiency
(although DLA1228-113g shows near-thermal excitation of the J=3 rotational level and a high SFR surface density). It thus appears likely that, similar to DLA1228-113g \citep{Kaur24}, DLA1020+2733g may also be a cold, rotating-disk galaxy. The high CO(3--2) integrated flux density implies that it should be straightforward to test this hypothesis via a high-resolution ALMA mapping study.

\section{Summary} \label{sec:sum}

We have used the VLA and ALMA to identify, via its CO(1--0) and CO(3--2) emission, a $z = 2.3568$ \hi-selected galaxy in the field of the $z = 2.3553$ DLA towards QSO~J1020+2733, at an impact parameter of $\approx 16$~kpc. The galaxy, DLA1020+2733g, has a molecular gas mass of $\rm M_{mol} = (2.84 \pm 0.42) \times 10^{11} \times (\alpha_{CO}/4.36) \, M_\odot$, the largest measured so far in an \hi-selected galaxy. The CO(3--2) and CO(1--0) lines are extended over a velocity range of $\approx 700$~\kms, in remarkable agreement with the extent of the low-ionization metal absorption lines detected in our Keck ESI spectrum towards the QSO. The velocity spread of the metal absorption lines is $\Delta V_{\rm 90} = 480 \pm 50$~\kms, one of the highest ever measured in a DLA. The ratio of CO(3--2) and CO(1--0) line luminosities yields an excitation of $r_{13} = 0.513 \pm 0.081$ for the J=3 rotational level; this is the first case of clear sub-thermal excitation of the J=3 level for a massive galaxy at $z >2$, suggesting that the DLA galaxy is likely to have a low SFR surface density. The ALMA-detected rest-frame 349-GHz dust continuum yields an SFR of $\lesssim 400$~M$_\odot$~yr$^{-1}$ for DLA1020+2733g, and a molecular gas depletion timescale of $\gtrsim 0.6$~Gyr. This is longer than the typical molecular gas depletion timescales in main-sequence galaxies at $z \approx 2.3$, suggesting that DLA1020+2733g has a relatively low star-formation efficiency for its molecular gas mass. We obtain an \hi\ column density of $(5.1 \pm 1.7) \times 10^{20}$~\cm\ and a metallicity of [Z/H]=$+0.28 \pm 0.16$ for the $z = 2.3553$ DLA, continuing the trend of the highest-metallicity DLAs arising in or around massive galaxies at $z \approx 2$. The sub-thermal CO excitation, large velocity spread, and long gas depletion timescale suggest that DLA1020+2733g is likely to be a cold, rotating disk galaxy.

\begin{acknowledgements}
We thank an anonymous referee for useful  comments on an earlier draft of this paper. BK and NK acknowledge support from the Department of Atomic Energy, under project 12-R\&D-TFR-5.02-0700; NK also acknowledges support from the Science and Engineering Research Board of the Department of Science and Technology via a J. C. Bose Fellowship (JCB/2023/000030). This material is based upon work supported by the National Science Foundation under Grant No. 2107989. ALMA is a partnership of ESO (representing its member states), NSF (USA) and NINS (Japan), together with NRC (Canada), NSC and ASIAA (Taiwan), and KASI (Republic of Korea), in cooperation with the Republic of Chile. The Joint ALMA Observatory is operated by ESO, AUI/NRAO and NAOJ. The National Radio Astronomy Observatory is a facility of the National Science Foundation operated under cooperative agreement by Associated Universities, Inc. The data reported in this paper are available through the ALMA archive (https://almascience.nrao.edu/alma-data/archive) with project code: ADS/JAO.ALMA \#2023.1.01415.S. Some of the data presented herein were obtained at Keck Observatory, which is a private 501(c)3 non-profit organization operated as a scientific partnership among the California Institute of Technology, the University of California, and the National Aeronautics and Space Administration. The Observatory was made possible by the generous financial support of the W. M. Keck Foundation.
\end{acknowledgements}

\facilities{ALMA, VLA, Keck:II(ESI).}

\software{{\sc casa} \citep{casa22}; {\sc gdl} \citep{Coulais10}; {\sc calR} \citep{Chowdhury21}.}
%https://ui.adsabs.harvard.edu/abs/2010ASPC..434..187C/abstract
\bibliography{bibliography}{}
\bibliographystyle{aasjournal}

\end{document}